# Performance comparison of an AI-based Adaptive Learning System in China

Wei Cui  Zhen Xue  and  Khanh-Phuong Thai
Yixue Squirrel AI Learning Inc., Shanghai, China

cuiwei@songshuai.com, xuezhen@mail.ustc.edu.cn, kp@songshuai.com

*Abstract*—adaptive learning systems stand apart from traditional learning systems by offering a personalized learning experience to students according to their different knowledge states. Adaptive systems collect and analyse students' behavior data, update learner profiles, then accordingly provide timely individualized feedback to each student. Such interactions between the learning system and students can improve the engagement of students and the efficiency of learning. This paper evaluates the effectiveness of an adaptive learning system, "Yixue Squirrel AI" (or Yixue), on English and math learning in middle school. The effectiveness of the Yixue's math and English learning systems is respectively compared against (1) traditional classroom math instruction conducted by expert human teachers and (2) BOXFiSH, another adaptive learning platform for English language learning. Results suggest that students achieved better performance using Yixue adaptive learning system than both traditional classroom instruction by expert teachers and another adaptive learning platform.

## I. Introduction

The widespread availability of computers and network connections inside and outside schools has drawn greater attention to technology-based learning systems. Studies have shown the effectiveness of such systems (e.g., VanLehn, 2011) and can promote student engagement (e.g., Kuh, 2003). The analysis on 74 studies showed that, educational technology intervention produce positive impacts (Cheung & Slavin, 2013; Steenbergen-Hu & Cooper, 2013, 2014). One major subset of such technology-based interventions is adaptive learning systems.

Through machine learning algorithms and data analytics techniques, adaptive learning systems stand apart from traditional learning systems by dynamically offering personalized learning experiences to students according to their different knowledge states. The intent is to determine what a student really knows and to move the student through a sequential path to certain learning goals. Learning products with adaptive features, such as Cognitive Tutors®, i-Ready®, Achieve3000®, Knewton®, RealizIt®, and ALEKS®, DreamBox® Learning, collect and analyze learners' behavior data, update learner profiles, and accordingly provide timely individualized feedback to each learner. As students spend more time in such a system, it can better estimate their abilities and can personalize instruction to best fit their strengths and weaknesses (e.g., van Seters et al., 2012).

As such adaptive learning systems are becoming more prevalent, there is an increased need in evaluating their effectiveness at helping students learn. There have been few rigorous evaluation studies, but previous findings from the United States suggest positive results. An meta-analysis on learning data including 1,600 adaptive courses and 4,800 non-adaptive courses showed that the adaptive courses were more effective in improving student performance (Bomash & Kish, 2015). Pane, Griffin, McCaffrey, & Karam (2014) conducted a large-scale effectiveness study of Cognitive Tutor Algebra I, a representative intelligent tutoring system, in diverse real-world school contexts and found promising positive impact of Cognitive Tutor after two years over traditional classroom instruction.

Even fewer experimental studies on the impact of adaptive learning systems have been done in China because the development of adaptive learning systems is still in the early stage in China, although online education in China has developed rapidly in recent years. The 42nd China statistical report on internet development in China shows that, by June 2018 the number of online education users in China had reached 172 million, accounting for 21.4% of total Internet users.

The purpose of this research is to further evaluate the impact of one such intelligent adaptive learning system, Yixue Squirrel AI (abbreviated as Yixue in this paper) on students' math and English language learning. Yixue's efficacy has been reported previously in 2018 International Conference on Computer Supported Education (Li, Cui, Xu, Zhu, & Feng, 2018) and 2018 Conference on Artificial Intelligence in Education (Feng, Cui, & Wang, 2018). This research looks to illustrate the automatic adaptive mechanism of Yixue, and confirm earlier findings by comparing Yixue against classroom instruction by expert teachers (Study 1) and another learning platform (Study 2).

## II. Yixue Intelligent Adaptive Learning System

The Yixue adaptive learning system was launched in 2016 and presently has over 100,000 users. Figure 1 shows the working mechanism of Yixue. Students only interact with the contents which are delivered to the students through a computer user interface, while the other parts of the figure work in the background server, not seen by the students.

The interaction between the learner and the content constantly produces learner data with time-stamp. The data, together with learner historical information data, are in real time sent to learning analytics engine, where the student profile or ability on each skill is updated by mining and analysis of the learning process.

As shown in Figure 1, the analysis results are sent to the intelligent adaptive learning engine to be diagnosed. In the intelligent adaptive learning engine, fine-grained knowledge map, together with knowledge space theory and information theory, are implemented to precisely diagnose the student's knowledge state with as few questions as possible. Accordingly, the engine in real time aims to dynamically determine the most optimal learning path so that it can deliver appropriate learning content for maximal learning effect.

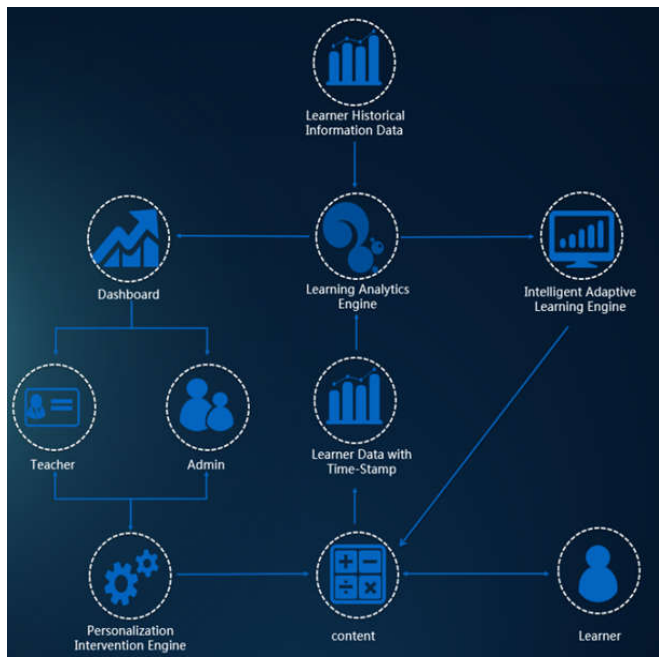

Figure 1. Working mechanism of Yixue

As shown in Figure 1, results from the learning analytics engine are also sent to the dashboard to be graphically displayed. Teacher can check the dashboard in real time to monitor students' learning progress, such as the student's answers and their response duration for each question, so the teacher can identify individual students who might be struggling, what she is struggling with, and provide her with corresponding support during class time. Teacher can instruct the student face-to-face if necessary. An administrator, such as a schoolmaster can also check the dashboard, to review student progress on each lesson and to view their score summary.

When a student starts a lesson in Yixue, she receives a diagnostic pretest to evaluate her mastery of the knowledge points associated with that lesson's learning targets. The subsequent learning phase focuses on the knowledge points she did not "master". The mastery criteria for our products are different from each other. In the math course product used in our study, the mastery criterion refers to the ability value scores above 0.75 on the knowledge point. In the English course product used in our study, the mastery criterion is scores above 0.85 on the knowledge point.

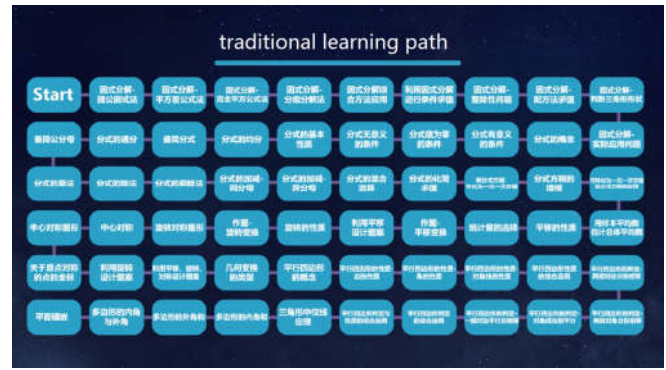

(a)

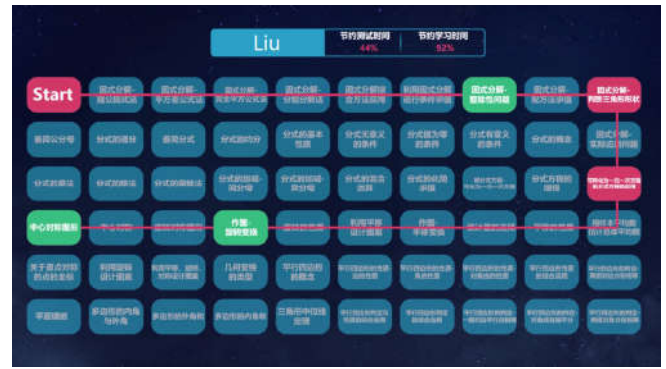

(b)

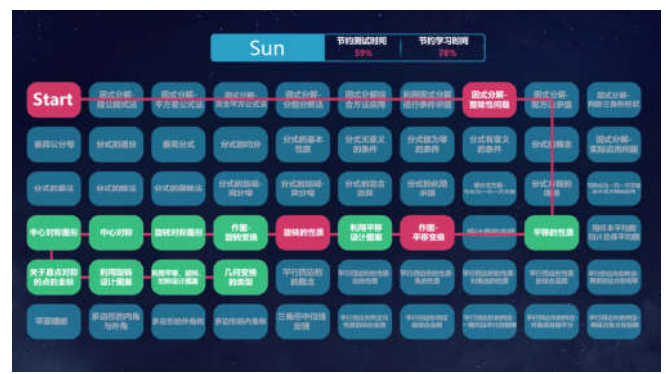

(c)

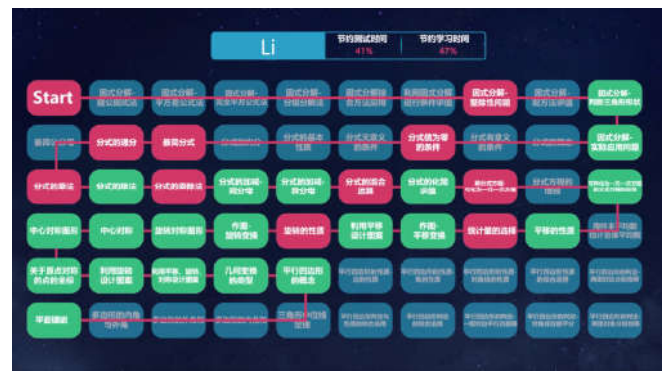

(d)

Figure 2. Sample personalized learning paths

Yixue aims to mimic an expert personal tutor. It evaluates the student's knowledge state and ability value by diagnosing the learning process, and then accordingly deliver a personalized learning path with appropriate learning materials for the student.

Figure 2 is an illustration of personalized learning paths. Figure 2(a) is a traditional linear learning path, in which blue boxes represent different knowledge points, and all students are generally expected to learn the knowledge points one by one through the path shown by the purple lines linking the boxes. In Figure 2(b)(c)(d), the red and green boxes linked by red lines represent weak knowledge points that should be learned by the student, according to the diagnosis in the pre-test. The dark blue boxes are the knowledge points skipped because the student has demonstrated mastery on those in the pre-test. The green boxes represent the knowledge points mastered after this session of learning, while the red ones are not yet mastered after this session of learning and still need to be learned. Unlike traditional learning path, the learning path in Yixue for each student is different, focusing on learning weak knowledge points and skipping those knowledge points already mastered. As shown in Figures 2(b), (c), and (d), student Liu has the fewest weak knowledge points to learn and student Li has the most weak knowledge points to learn. Through this personalized learning path, student Liu would save 92% of learning time compared to traditional learning path, Sun would save 78% of learning time, and Li would save 47%.

## III. RESEARCH AIM

We conducted two studies to compare the effectiveness of (1) Yixue math middle school program against business-as-usual classroom instruction and (2) Yixue English middle school program against BOXFiSH online learning product. The primary purpose of the studies was to further evaluate the impact Yixue has on improving student learning and to determine the likely magnitude of those effects, if they exist. In earlier studies, Feng et al. (2018) compared Yixue with classroom instruction by expert human teachers, and found that Yixue on average scored 8.56 points higher than instruction by human teachers for grade 8 students with a statistically significant difference ($F(1, 32) = 3.35$, $p = 0.08$, $r^2 = 53.08$) and an effect size of $g = 0.48$，and Yixue scored 3.41 points higher than instruction by human teachers for grade 9, without a statistically significant difference ($F(1, 40) = 2.02$, $p = .16$, $r^2 = 49.38$) but a substantial effect size for the Yixue intervention, $g = 0.32$. In another study, Li et al. (2018) compared Yixue with Magic Grid, another popular online learning platform, and found that students who used Yixue on average scored 3.8 points higher on the post-test than students who used Magic Grid, after controlling for differences in pre-test scores, and the effect was marginally significant ($F(2, 84) = 104.6$, $p = .09$, $r^2 = 0.71$).

## IV. METHODS

*Experiment 1: Comparison of Yixue and Classroom Instruction by Expert Teachers*

**Sample**.

Typical middle school students, 13 to 15 years old, were sampled from Chengdu. Students were randomly assigned into the control group and treatment group using stratified block randomization (Trochim, Donnelly, & Arora, 2016, p. 229). A total of 203 students took part in the experiment, of whom 101 students were assigned to the treatment group and 102 students were assigned to the control group.

90 students of the treatment group finished the experiment, where Yixue system was used as the intervention. 73 students of the control group finished the experiment, where they were instructed by expert teachers. Students who did not complete the study chose to drop out on their own volition. Following the pre-test, students in the control group were then split into three sub-groups to receive instructions by three teachers. The teachers had taught math in local middle or high school for 8 to 18 years.

**Experiment procedure**.

Table 1. Schedule of Experiment 1

| 2018-04-29 | |
|---|---|
| 13:00-13:05 | Introduction |
| 13:05-13:10 | Questionnaire |
| 13:10-14:00 | Pre-test |
| 14:10-15:00 | Instruction by Yixue/human teachers |
| 2018-04-30 | |
| 13:00-13:50 | Instruction by Yixue/human teachers |
| 14:00-14:50 | Instruction by Yixue/human teachers |
| 15:00-15:50 | Instruction by Yixue/human teachers |
| 2018-05-01 | |
| 13:00-13:50 | Instruction by Yixue/human teachers |
| 14:00-14:50 | Instruction by Yixue/human teachers |
| 15:00-15:50 | Post-test |
| 15:50-16:00 | Questionnaire |

The study lasted for 3 days during a national vacation. The schedule of this study is shown in Table 1, which is the same for every student, except the treatment group received Yixue and the control group received teacher instruction. In the introduction on the first day, students received explanation of the experimental procedure. The intervals between the activities listed in Table 1 are rest breaks. The learning contents covered by both groups included the Pythagorean theorem and its application, real numbers, triangles, integer expression, properties of a triangle, and reflection symmetry. These contents had already been instructed during the semester prior to our study. Students in the treatment group used Yixue system and worked on topics above without assistance from teacher. In the control group, teachers taught the topics according to local learning standards. Students in the control group were not supported by any online learning programs. The questionnaires at the beginning and end of the study included demographic questions and students' ratings of their learning experience during the study.

**Pre- and post-test data.**

Items in the pre- and post-tests were constructed by an experienced teacher in a local school (not part of the research team). When developing the tests, the teacher was only provided the test topics and the associated learning standards. Two independent, experienced subject matter experts reviewed the pre- and post-tests to ensure that they were comparable in their coverage, overall difficulty, types of items, and alignment with the learning standards. The experts also checked to make sure that the test items were not over-aligned with Yixue learning content. Each test composed of 30 multiple-choice, completion, calculation, and application items with a total score of 100 points. The tests are scored by teachers, then scanned and sent to the third-party agency to be reviewed.

**Analysis and results**.

Figure 3 shows the mean gain scores from pre-test to post-test from the treatment group and the control group. While both treatment and control groups showed improvements from pre-test to post-test, those who used Yixue showed 4.19 times greater gains that those who received traditional classroom instruction (Treatment gain $M$ = 9.38, $SD$ = 11.08 vs. Control gain $M$ = 1.81, $SD$ = 10.91, Hedges' $g$ = .68). A one-way analysis of covariance (ANCOVA) with pre-test score as a covariate confirmed this result, $F(1,160)$ = 16.80, $p < .001$, partial $\eta^2$ = .10.

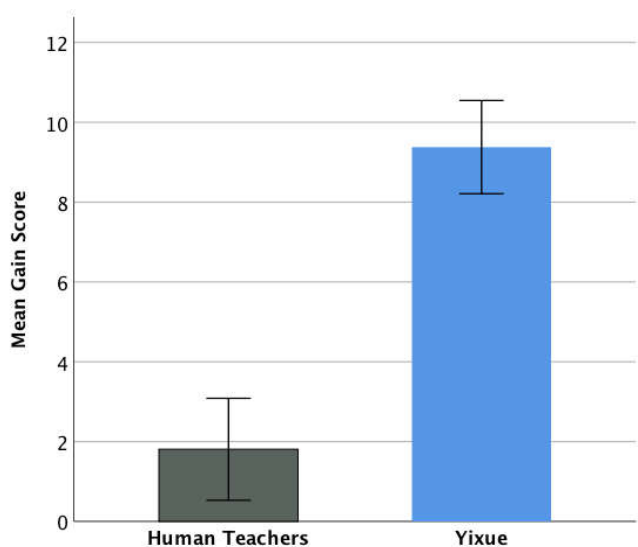

Figure 3. Growth from pre-test to post-test from students in the control group (human teachers) and the treatment group (Yixue). Error bars represent ± 1 standard error.

There was no statistically significant difference between treatment and control groups on their prior knowledge at pre-test, $t$(155.13, unequal variance assumed) = 1.49, $p$ = .14, Hedges' $g$ = .25. However, pre-test was also a strong predictor of learning gains, $F(1,160)$ = 6.14, $p$ = .014, partial $\eta^2$ = .10, with those who start lower at pre-test tended to show higher gains.

After the experiment, researchers and local teachers randomly interviewed 68 students who used Yixue adaptive learning system. The interviewees answered questions about their satisfaction about Yixue's learning contents, learning experience, ease of use, and effectiveness. Some of the interview results are shown in Figure 4. 87% of the interviewees had positive judgment of learning math using Yixue; 93% of the interviewees rated YiXue as adaptive, with 78% clearly felt that it was adaptive to himself or herself; of the learning contents, 87% of the interviewees thought that the contents were tailored to their needs, with 68% clearly thought so; of the multi-media contents, 12% of the interviewees like all of them, 46% like animation, 30% like lecturing video, 9% like the powerpoint, 2% like animation and Powerpoint, 2% like none of them.

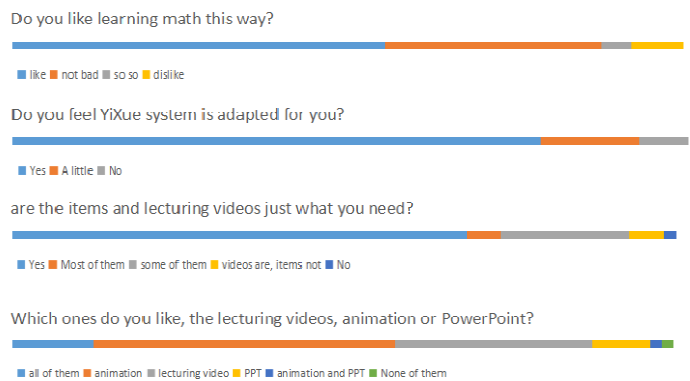

Figure 4. interview of Yixue users

*Experiment 2: Comparison of Yixue and BOXFiSH*

BOXFiSH® is an intelligent English-learning product with a curriculum system of varying difficulty levels for kindergarten, K12, TOELF and IELTS, based on fine-sorted knowledge points. BOXFiSH programs the learning process according to students' characteristics and aim to maximize learning efficiency through the instruction of AI and well-trained teachers domestic and abroad, transitioning students from passive learning to active learning. BOXFiSH provides a language environment through high-quality pictures, native foreign language, and interesting videos, visualizing the implication of knowledge points. It manufactured 9694 class hours of K12 curriculum, so that the students can re-learn the 12 types of 5402 knowledge points in their syllabi through a new and interesting way (BOXFiSH, 2018). BOXFiSH can be used in conjunction with K12 school English teaching, assisting teachers in flipped classroom. BOXFiSH is usually used together with classroom teaching. In this experiment, students in the control group studied for 2 class hours in classroom teaching, then 1 class hour using BOXFiSH, and so on.

For personalized learning, BOXFiSH can provide:

1. personalized delivery according to: the grade of student, the textbook edition and version the student is using, starting difficulty she prefers, time duration she has for learning, etc.

2. Intelligent adjustment of instruction in accordance with students' mastering of knowledge points by:

(1) prioritizing knowledge points of higher difficulty,

(2) continuing delivering related content until the knowledge point is mastered, and

(3) providing more difficult content after student finishes related content of the same difficulty.

**Sample.**

In the experiment, junior middle school students, 13 to 15 years old, were sampled from Zhengzhou Erqi District, Zhongyuan District, and Jinshui District. The sampling method is the same as experiment 1: stratified block randomization (Trochim, Donnelly, & Arora, 2016, p. 229). A total of 140 students participated in the experiment, of whom 70 students were assigned to the treatment group and 70 students were assigned to the control group.

Of all the participating students, 104 completed all steps (pre-test, learning, and post-test) of the experiment and finished the tests within given time. 46 students are of the treatment group who used Yixue, 58 students are of the control group who used BOXFiSH. All students who dropped out did so that their own volition.

**Experiment procedure.**

The procedure of experiment 2 was similar to that of experiment 1. Experiment 2 lasted two days. On the first day of the study, students in both groups took a paper-and-pencil pre-test on the topics they would review, "Grammar," which had already been instructed during their semester prior to this study. The grammar curriculum includes 18 topics, such as *wh*-questions beginning with "what", and the simple present tense. The BOXFiSH curriculum contained 200 multiple-choice questions. Students in the treatment group studied the the topics using Yixue on computer, and the control group studied with teachers and BOXFiSH. Students in the control group could ask teachers for help while using BOXFiSH. The learning schedule for both groups was identical, including the break time between online classes and the number of breaks. On the second day, at the end of the learning sessions, a paper-and-pencil post-test was administered to both groups. Students were given 25 minutes to finish the pre and post-tests.

**Pre- and post-test data.**

Items for pre- and post-tests were constructed and scored in the same way as in Experiment 1, except that in experiment 2, pre- and post-test were both composed of 30 multiple choice, completion, and translation questions.

**Analysis and results.**

Figure 5 shows the mean gain scores from pre-test to post-test from the treatment group and the control group. Similar to experiment 1, while both treatment and control groups showed overall improvements from pre-test to post-test, those who used Yixue achieved 4.62 times greater gains that those who used BOXFiSH in the same period (Yixue gain $M = 5.86$, $SD = 10.81$ vs. BOXFiSH gain $M = 1.04$, $SD = 8.93$). Because participating students came from three different school districts, we confirmed this finding using a multiple linear regression analysis on the gain score, controlling for school location and pre-test, $F(3, 100) = 30.90$, $p < .001$, $r^2 = .47$. Students in the treatment group had statistically greater gains than the control group controlling for their pretest score and school location, $\beta = .22$, $p = .003$, *Hedges' g* $= .49$. There was no statistically significant difference between treatment and control groups on their prior knowledge at pre-test, $t(101.34$, unequal variance assumed$) = .68$, $p = .50$, *Hedges' g* $= .13$. However, those with lower pre-test scores tended to show higher gains at post-test, $\beta = -.38$, $p < .001$. School was also a strong predictor of learning gains, $\beta = -.42$, $p < .001$.

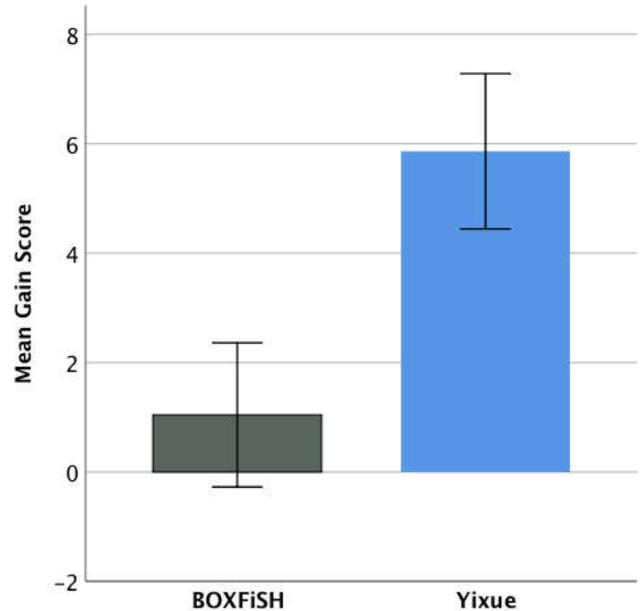

Figure 5. Gains from pre-test to post-test by students in the control group (BOXFiSH) and the treatment group (Yixue). Error bars represent ± 1 standard error.

V. CONCLUSION

In this paper, we introduced the automated adaptivity of the Yixue Squirrel AI learning system and its implementation model. The two studies in this paper showed that the Yixue adaptive learning system produced greater learning gains than (1) classroom instruction by expert human teachers for math and (2) BOXFiSH, an AI-based adaptive learning competitor for English, likely due to Yixue's fine granularity of knowledge points and intelligent adaptivity.

Because these studies were conducted during just a few days, future research will include a longer term study for a better evaluation of learning with Yixue over time, and with bigger sample size for higher precision. Future studies will also focus on establishing a more detailed user profile using additional qualitative data from student questionnaires and interviews.

The number of learning systems is increasing in China and around the world. However, there remains little in-depth investigation of learning results from such systems in the US as well as in China. With schools in China starting to introduce educational technologies into the classrooms, there is broad interest in understanding how to select and use such systems, which may also lead to improvement of the design of such systems. This paper presents the much-needed knowledge about adaptive learning in K‒12 instruction in China and around the world.